\documentclass[12pt]{article}
\usepackage{amscd,amsmath,amsxtra,amssymb,amstext,latexsym,amsthm}
\usepackage{graphicx}
\newcommand {\dl}   {\delta}       \newcommand {\e }  {\epsilon}

\newcommand {\lm}   {\lambda}      
          
\newcommand {\s }   {\sigma}       
         
\newcommand {\vf }  {\varphi}      
         \newcommand {\om}  {\omega}

\newcommand {\pl}   {\partial}     \newcommand {\nb}  {\nabla}
       
\newcommand   {\ex}{{\sf\,e}}            
\newcommand   {\re}{{\sf\,re\,}}         
\newcommand   {\const}{{\sf\,const}}     
\newcommand {\vol}  {\sqrt{{\scriptstyle |}g{\scriptstyle |}}}
   
\newcommand {\MO}  {{\mathbb O}}   
   \newcommand {\MR}  {{\mathbb R}}
\newcommand {\MS}  {{\mathbb S}}   

   \newcommand {\St}  {{\textsc{t}}}

\begin{document}
\title     {Torsional Elastic Waves in Double Wall Tube}
\author    {M. O. Katanaev
            \thanks{E-mail: katanaev@mi.ras.ru}\\ \\
            \sl Steklov Mathematical Institute,\\
            \sl ul.~Gubkina, 8, Moscow, 119991, Russia}
\date      {03 March 2015}
\maketitle
\begin{abstract}
We describe the double wall tube with cylindrical dislocation in the framework
of the geometric theory of defects. The induced metric is found. The dispersion
relation is obtained for the propagation of torsional elastic waves in the
double wall tube.
\end{abstract}
\section{Introduction}
Ideal crystals are absent in nature, and most of their physical properties, such
as plasticity, melting, growth, etc., are defined by defects of the crystalline
structure. Therefore, a study of defects is a topical scientific question of
importance for applications in the first place. At present, a fundamental theory
of defects is absent in spite of the existence of dozens of monographs
and thousands of articles.

One of the most promising approaches to the theory of defects is based on
Riemann--Cartan geometry, which involves nontrivial metric and torsion.
In this approach, a crystal is considered as a continuous elastic medium with
a spin structure. If the displacement vector field is a smooth function, then
there are only elastic stresses corresponding to diffeomorphisms of the
Euclidean space. If the displacement vector field has discontinuities, then
we are saying that there are defects in the elastic structure. Defects in the
elastic structure are called dislocations and lead to the appearance
of nontrivial geometry. Precisely, they correspond to a nonzero torsion tensor,
equal to the surface density of the Burgers vector. Defects in the spin
structure are called disclinations. They correspond to nonzero curvature tensor,
curvature tensor being the surface density of the Frank vector.

The idea to relate torsion to dislocations appeared in the 1950s [1--4].
\nocite{Kondo52,Nye53,BiBuSm55,Kroner58}
This approach is still being successfully developed (note reviews [5--11]),
\nocite{SedBer67,Kleman80A,Kroner81,DzyVol88,KadEde83,KunKun86,Kleine89}
and is often called the gauge theory of dislocations.

Some time ago we proposed the geometrical theory of defects [12--14]
\nocite{KatVol92,KatVol99,Katana05}. Our approach is essentially different from
others in two respects. Firstly, we do not have the displacement and rotational
angle vector fields as independent variables because, in general,
they are not continuous. Instead, the triad field and $\MS\MO(3)$-connection are
considered as independent variables. If defects are absent, then the triad and
$\MS\MO(3)$-connection reduce to partial derivatives of the displacement and
rotational angle vector fields. In this case the latter can be reconstructed.
Secondly, the set of equilibrium equations is different. We proposed purely
geometric set which coincides with that of Euclidean three dimensional gravity
with torsion. The nonlinear elasticity equations and principal chiral
$\MS\MO(3)$ model for the spin structure enter the model through the elastic and
Lorentz gauge conditions [14--16] \nocite{Katana03,Katana04,Katana05} which
allow to reconstruct the displacement and rotational angle vector fields
in the absence of dislocations in full agreement with classical models.

The advantage of the geometric theory of defects is that it allows one to
describe single defects as well as their continuous distributions.

In the present paper, we consider propagation of torsional elastic waves in
double wall tube with cylindrical dislocation. This defect was first described
in \cite{deBKat09}. The Schr\"odinger equation for the double wall tube was
solved in \cite{deKaKoSh10} and applied to double wall nanotubes. A similar
problem was also solved for the cylindrical waveguide with wedge dislocation
\cite{Katana15B}.
\subsection{Double wall tube}
Let us describe double wall tube with cylindrical dislocation in the framework
of the geometric theory of defects.

\begin{figure}[h,t]
\hfill\includegraphics[width=.4\textwidth]{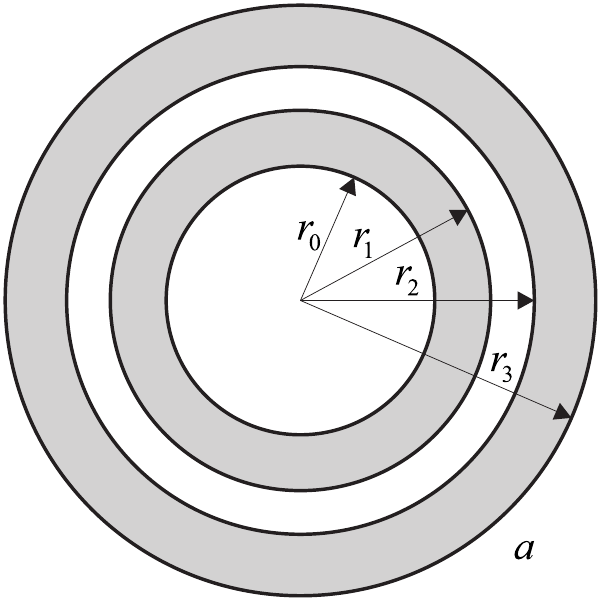}
\hfill  \hspace*{.06\textwidth} \hfill
\includegraphics[width=.4\textwidth]{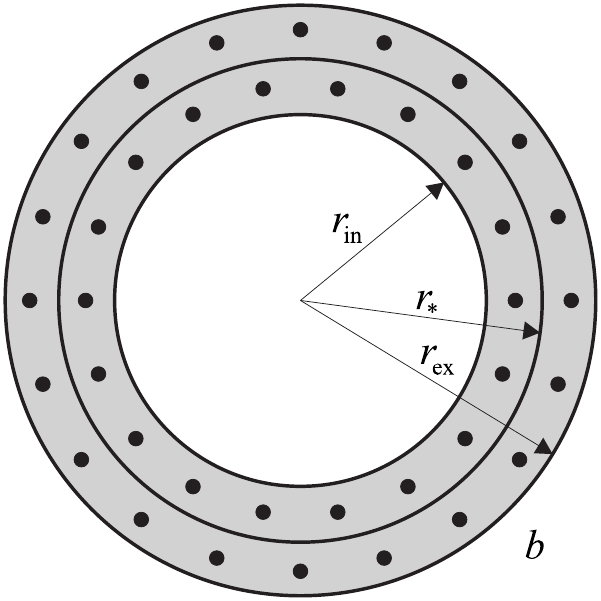}
\centering \caption{\label{ftubed} Section $z=\const$ of double wall tube before
(\textit{a}) and after (\textit{b}) defect creation.}
\end{figure}

We consider cylindrical coordinates
$\lbrace x^\mu\rbrace=\lbrace r,\vf,z\rbrace$, $\mu=1,2,3$ in tree dimensional
Euclidean space  $\MR^3$. Let there be two thick tubes $r_0\le r\le r_1$ and
$r_2\le r\le r_3$ of elastic media, each axis coinciding with the $z$ axis.
We suppose that $r_0<r_1<r_2<r_3$ (see Fig.\ref{ftubed},\textit{a}, where
a section $z=\const$ is shown).
Now we make one tube with the inside cylindrical dislocation in the following
manner. We stretch symmetrically the inner tube and compress the outer one. Then
glue together the external surface of the inner tube with the internal surface
of the outer tube. Afterwards the media comes to some equilibrium state. Due
to rotational and translational symmetry we obtain one tube
$r_\text{in}\le r\le r_\text{ex}$ with the axis which coincides with the $z$
axis (see Fig.\ref{ftubed},\textit{b}). Radii of cylinders constituting tube
surfaces are mapped as follows
\begin{equation*}
  r_0\mapsto r_\text{in},\qquad r_1,r_2\mapsto r_*,\qquad
  r_3\mapsto r_\text{ex}.
\end{equation*}
The gluing is performed along the cylinder $r_*$, and there is cylindrical
defect (dislocation) because part of the media between tubes is removed.

The obtained double wall tube with cylindrical dislocation is rotationally and
translationally symmetric.

The constructed model of the tube with cylindrical dislocation can be considered
as continuous model of double wall nanotube (for a general review, see
\cite{Ebbese97,SaDrDr98,Harris99}). Consider double wall nanotube having two
atomic layers. Suppose the inner layer has 18 and outer layer has 20 atoms which
are shown in Fig.\ref{ftubed},\textit{b} by points. Natural length measure here
is the interatomic distance. Then the length of a circle has a jump when one
goes from inner to outer layer. In the geometric theory of defects, it means
that the metric component $g_{\vf\vf}$ is not continuous in cylindrical
coordinates. The corresponding model will be described below.

To find radii $r_\text{in}$, $r_*$, and $r_\text{ex}$ we have to solve the
classical elasticity problem.

Let us define the displacement vector field by $u^i(x)$, $i=1,2,3$,
\begin{equation}                                                  \label{qcasew}
  y^i\mapsto x^i=y^i+u^i(x),
\end{equation}
where $y^i$ and $x^i$ are coordinates of a point before and after deformation
respectively. We consider the displacement field as a vector function on points
of media after deformation and gluing. This is more adequate because the
resulting media after gluing is a connected manifold (before the gluing
procedure, each tube represents a connected component). In equilibrium state,
the vector displacement field
satisfies equation
\begin{equation}                                                  \label{qbhfyk}
  (1-2\s)\triangle u_i+\pl_i\pl_j u^j=0,
\end{equation}
where $\s$ is the Poisson ratio and $\triangle$ is the Laplacian. For
convenience, we consider components of the displacement vector field with
respect to the orthonormal basis
\begin{equation*}
  u=u^{\hat r}e_{\hat r}+u^{\hat\vf}e_{\hat\vf}+u^{\hat z}e_{\hat z},
\end{equation*}
where
\begin{equation*}
  e_{\hat r}=\pl_r,\qquad e_{\hat\vf}=\frac1r\pl_\vf,\qquad e_{\hat z}=\pl_z.
\end{equation*}
We denote indices with respect to the orthonormal basis by hat:
\begin{equation*}
  \lbrace i\rbrace=\lbrace \hat r,\hat \vf,\hat z\rbrace,\qquad
  \lbrace \mu\rbrace=\lbrace r,\vf,z\rbrace.
\end{equation*}
The Latin indices referred to an orthonormal basis are rased and lowered by
Kronecker symbol: $u_i:=u^j\dl_{ji}$.

The divergence and Laplacian have the following form in cylindrical coordinates
\begin{equation}                                                  \label{qbndki}
\begin{split}
  \nb_i u^i&=\frac1r\pl_r(ru^{\hat r})
  +\frac1r\pl_\vf u^{\hat\vf}+\pl_z u^{\hat z},
\\
  \triangle u_{\hat r}&=\frac1r\pl_r(r\pl_r u_{\hat r})
  +\frac1{r^2}\pl^2_{\vf\vf}u_{\hat r}+\pl^2_{zz}u_{\hat r}
  -\frac1{r^2}u_{\hat r}-\frac2{r^2}\pl_\vf u_{\hat \vf},
\\
  \triangle u_{\hat \vf}&=\frac1r\pl_r(r\pl_r u_{\hat\vf})
  +\frac1{r^2}\pl^2_{\vf\vf}u_{\hat\vf}+\pl^2_{zz}u_{\hat\vf}
  -\frac1{r^2}u_{\hat\vf}+\frac2{r^2}\pl_\vf u_{\hat r},
\\
  \triangle u_{\hat z}&=\frac1r\pl_r(r\pl_r u_{\hat z})
  +\frac1{r^2}\pl^2_{\vf\vf}u_{\hat z}+\pl^2_{zz}u_{\hat z}.
\end{split}
\end{equation}

From the symmetry of the problem, we deduce that only radial component of the
displacement field differs from zero, and it does not depend on the angle $\vf$
and $z$ coordinates:
\begin{equation*}
  \lbrace u^i\rbrace=\lbrace u^{\hat r}:=u(r), u^{\hat\vf}=0,u^{\hat z}=0
  \rbrace.
\end{equation*}
Equation (\ref{qbhfyk}) for zero $u_{\hat\vf}$ and $u_{\hat z}$ components are
automatically satisfied. It is easy to check that the radial derivative of the
divergence,
\begin{equation*}
  \pl_{\hat r}\pl_ju^j=\pl_r\left(\frac1r\pl_r(ru)\right)
  =\pl^2_{rr}u+\frac1r\pl_r u-\frac1{r^2}u,
\end{equation*}
coincides with the Laplacian
\begin{equation*}
  \triangle u_{\hat r}=\frac1r\pl_r(r\pl_r u)-\frac1{r^2}u
  =\pl^2_{rr}u+\frac1r\pl_r u-\frac1{r^2}u.
\end{equation*}
Therefore the radial component of Eq.(\ref{qbhfyk}) takes the form
\begin{equation}                                                  \label{qbncgd}
  \pl_r\left(\frac1r\pl_r(ru)\right)=0.
\end{equation}
A general solution of this equation depends on two integration constants:
\begin{equation*}
  u=c_1r+\frac{c_2}r,\qquad c_{1,2}=\const.
\end{equation*}

Note that the equilibrium equation (\ref{qbncgd}) does not depend on the
Poisson ratio $\s$. This means that the cylindrical dislocation is the
geometrical defect.

Boundary conditions have to be imposed to fix the integration constants.
Let us introduce notation for inner and outer tubes:
\begin{equation*}
  u=\begin{cases} u_\text{in}, & \quad r_\text{in}\le r\le r_*, \\
  u_\text{ex}, & \quad r_*\le r\le r_\text{ex}.
\end{cases}
\end{equation*}
Now boundary conditions are to be imposed. We assume that the surface of two
wall nanotube is free, i.e.\ the deformation tensor is zero on the boundary:
\begin{equation}                                                  \label{qnbdtg}
  \left.\frac{du_\text{in}}{dr}\right|_{r=r_\text{in}}=0,\qquad
  \left.\frac{du_\text{ex}}{dr}\right|_{r=r_\text{ex}}=0.
\end{equation}
We assume also that the media is in equilibrium. It means that elastic forces on
the gluing surface must be equal to zero:
\begin{equation}                                                  \label{qbdjii}
  \left.\frac{du_\text{in}}{dr}\right|_{r=r_*}=
  \left.\frac{du_\text{ex}}{dr}\right|_{r=r_*}.
\end{equation}

Each of boundary conditions (\ref{qnbdtg}) defines one integration constant for
internal and external solutions:
\begin{equation}                                                  \label{qbndtr}
\begin{split}
  u_\text{in}&=~~a\left(r+\frac{r^2_\text{in}}r\right)>0,\qquad a=\const>0,
\\
  u_\text{ex}&=-b\left(\frac1r+\frac r{r_\text{ex}^2}\right)<0,\qquad
  b=\const>0.
\end{split}
\end{equation}
Signs of the integration constants $a$ and $b$ are chosen in such a way that
displacement vector is positive and negative for inner and outer tubes
respectively. This in agreement with the imposed problem.

Substitution of obtained solutions (\ref{qbndtr}) into the gluing condition
(\ref{qbdjii}) defines the ratio of integration constants:
\begin{equation}                                                  \label{qbkkos}
  r_*^2=r^2_\text{ex}\frac{ar^2_\text{in}+b}{ar^2_\text{ex}+b}\qquad
  \Leftrightarrow\qquad
  b=ar^2_\text{ex}\frac{r_*^2-r^2_\text{in}}{r^2_\text{ex}-r_*^2}.
\end{equation}

The entirety condition for the media is
\begin{equation}                                                  \label{qfaelk}
\begin{split}
  r_*&=r_1+a\left(r_*+\frac{r^2_\text{in}}{r_*}\right),
\\
  r_*&=r_2-b\left(\frac1{r_*}+\frac{r_*}{r^2_\text{ex}}\right).
\end{split}
\end{equation}
These equations allow to find the distance between initial tubes which
characterize the cylindrical dislocation:
\begin{equation}                                                  \label{qbhjsu}
  l:=r_2-r_1=2ar_*\frac{r^2_\text{ex}-r^2_\text{in}}{r^2_\text{ex}-r^2_*},
\end{equation}
where expression for $b$ (\ref{qbkkos}) is used. Afterwards we find the
integration constants:
\begin{equation}                                                  \label{qnhydh}
  a=\frac l{2r_*}\,\frac{r^2_\text{ex}-r^2_*}{r^2_\text{ex}-r^2_\text{in}},
  \qquad b=\frac{lr^2_\text{ex}}{2r_*}\,\frac{r^2_*-r^2_\text{in}}{r^2_\text{ex}
  -r^2_\text{in}}.
\end{equation}

\begin{figure}[h,t]
\hfill\includegraphics[width=.4\textwidth]{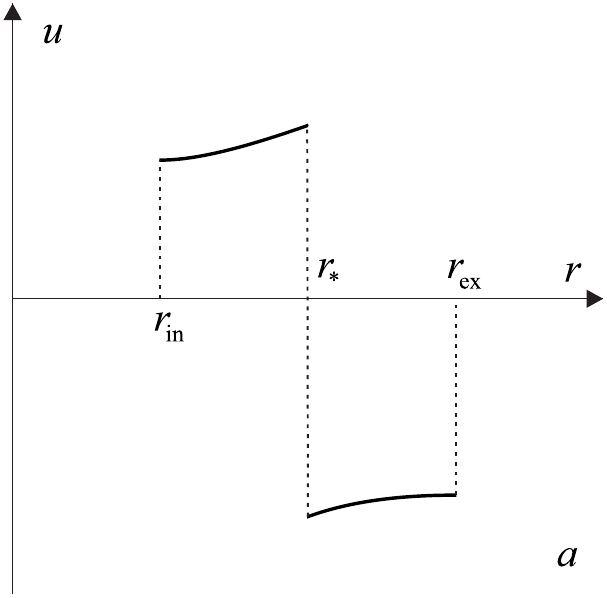}
\hfill  \hspace*{.06\textwidth} \hfill
\includegraphics[width=.4\textwidth]{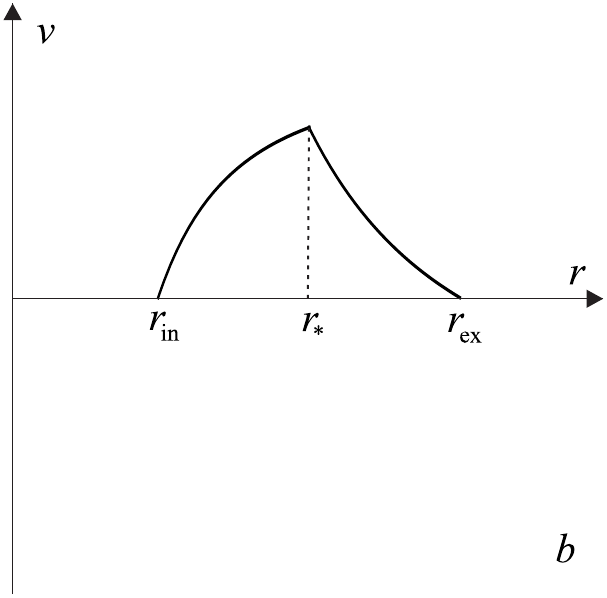}
\centering \caption{\label{ftubeu} Qualitative behaviour of the radial component
of the displacement vector field for the double wall tube (\textit{a}). The
derivative of the displacement vector field (\textit{b}).}
\end{figure}

Thus we find the displacement vector field
\begin{equation}                                                  \label{qbvxgt}
  u(r)=\begin{cases}
    ~~a\left(r+\frac{r^2_\text{in}}r\right)>0,\qquad &r_\text{in}\le r<r_*, \\
    -b\left(\frac1r+\frac r{r_\text{ex}^2}\right)<0,\qquad &r_*<r\le r_\text{ex}
\end{cases}
\end{equation}
for double wall tube where constants $a$ and $b$ are given by
Eqs.(\ref{qnhydh}). Qualitative behaviour of this vector field is shown in
Fig.\ref{ftubeu},\textit{a}. Differentiation of this vector field in domains
$r_\text{in}<r<r_*$, $r_*<r<r_\text{ex}$ and its extension to the point $r_*$
by continuity yields the function
\begin{equation}                                                  \label{qnjuht}
  v(r):=\frac{du}{dr}=\begin{cases} a\left(1-\frac{r_\text{in}^2}{r^2}\right)>0,
  \qquad &r_\text{in}\le r\le r_*, \\
  b\left(\frac1{r^2}-\frac1{r^2_\text{ex}}\right)>0,\qquad
  &r_*\le r\le r_\text{ex},
\end{cases}
\end{equation}
which is depicted in Fig.\ref{ftubeu},\textit{b}.

Note that media entirety condition (\ref{qfaelk}) leads to the jump of the
displacement vector field at the point $r_*$:
\begin{equation*}
  l:=r_2-r_1=u_\text{in}(r_*)-u_\text{ex}(r_*).
\end{equation*}
Since vector field $u$ has discontinuity at the point $r_*$, the formal
derivative of $u$ contains $\dl(r-r_*)$. This $\dl$-function is thrown away in
the geometric theory of defects \cite{Katana05}.

Double wall tube with cylindrical dislocation is parameterized by four constants
$r_0$, $r_1$, $r_2$, $r_3$ or $r_\text{in}$, $r_*$, $r_\text{ex}$, $l$.
Formulae (\ref{qnhydh}) define constants $a$, $b$ and, consequently, the
displacement vector field through the second set of parameters. It follows from
definition (\ref{qcasew}) that there is one-to-one correspondence between two
sets of parameters.

Now we calculate the metric induced in double wall tube. By definition, it has
the form
\begin{equation}                                                  \label{qbnctf}
  g_{\mu\nu}(x)=\frac{\pl y^\rho}{\pl x^\mu}\frac{\pl y^\s}{\pl x^\nu}
  \overset{\circ}g_{\rho\s}(y),
\end{equation}
where $\overset{\circ}g_{\rho\s}(y)$ is the Euclidean metric in cylindrical
coordinates. The relation between coordinates before and after defect creation
is given by equality (\ref{qcasew}). Using explicit form of displacement
vector field (\ref{qbvxgt}) we find the induced metric in the double wall tube
\begin{equation}                                                  \label{qbcnct}
  ds^2=(1-v)^2dr^2+(r-u)^2d\vf^2+dz^2.
\end{equation}
The component $g_{rr}=(1-v)^2$ of this metric is continuous function but its
derivative has a jump at $r_*$. The component $g_{\vf\vf}=(r-u)^2$ is
discontinuous at the point $r=r_*$.

The volum element for metric (\ref{qbcnct}) is
\begin{equation*}
  \vol=(1-v)(r-u).
\end{equation*}
The right hand side of this relation is positive because both multipliers are
positive. The second multiplier $r-u=y>0$ is positive by construction. The first
multiplier is also positive. Indeed, the function $v$ has the maximum at
$r=r_*$. At this point, the following inequality holds
\begin{equation*}
  v(r_*)=a\frac{r^2_*-r^2_\text{in}}{r^2_*}=\frac{r_*-r_1}{r_*}\,
  \frac{r^2_*-r^2_\text{in}}{r^2_*+r^2_\text{in}}<1,
\end{equation*}
where expression (\ref{qfaelk}) for $a$ is used.

The circumference is a geometric invariant. It is equal to
$2\pi\big(r-u(r)\big)$ for metric (\ref{qbnctf}). When we go from the inner tube
to the outer one it has the jump $2\pi l$ where $l$ is the distance between
tubes before the dislocation is made. This observation agrees with the
continuous model of double wall tube.
\subsection{Torsional waves}
Here we consider torsional waves in the double wall tube  with cylindrical
dislocation described in the previous section. We denote the displacement vector
field by the new letter $w$ because the letter $u$ was used in the previous
section for the displacements corresponding to the defect creation. The total
displacement vector field is equal to the sum $u+w$ where $u$ corresponds to
the defect creation and $w$ describes oscillations in the double wall tube with
cylindrical dislocation. By definition, the displacement vector field $w$
satisfies the wave equation
\begin{equation}                                                  \label{qnbhfy}
  \rho_0\ddot w_i-\mu\tilde\triangle w_i
  -(\lm+\mu)\tilde\nabla_i\tilde\nabla_jw^j=0,
\end{equation}
where the covariant derivative $\tilde\nb_i:=e^\mu{}_j\tilde\nb_\mu$ and
Laplace--Beltrami operator
$\tilde\triangle:=g^{\mu\nu}\tilde\nb_\mu\tilde\nb_\nu$ are defined by metric
(\ref{qbcnct}) of double wall tube.

This equation is covariant with respect to changing of coordinate systems. It
can be solved in cylindrical coordinates $r,\vf,z$ after the defect creation
with metric (\ref{qbcnct}). However it is easier to follow another way. We solve
the wave equation in cylindrical coordinates $y,\vf,z$ where $y$ denotes the
old radial coordinate before the defect creation
\begin{equation*}
  y:=r+u,
\end{equation*}
and afterwards impose necessary boundary conditions. It is easier because the
metric is Euclidean in the initial coordinate system.

Let us consider torsional waves. In this case only angular component of the
displacement vector field differs from zero:
\begin{equation*}
  \lbrace w^i\rbrace=\lbrace w^{\hat r}=0,w^{\hat\vf}=w^{\hat\vf}(t,r),
  w^{\hat z}=0\rbrace.
\end{equation*}
From symmetry consideration, the angular component $w^{\hat\vf}$ does not depend
on $\vf$ and $z$. For this vector field the $\hat r$ and $\hat z$ components
of Eq.(\ref{qnbhfy}) are automatically satisfied. It follows that the
dilation for torsional waves is equal to zero
\begin{equation*}
  \e:=\pl_iw^i=0,
\end{equation*}
and consequently torsional waves take place without media compression.

We look for solution of Eq.(\ref{qnbhfy}) in the plain wave form
\begin{equation}                                                  \label{qhdoph}
  w_{\hat\vf}=\re \left[W(y)\ex^{i(kz-\om t)}\right],
\end{equation}
where $W(y)$ is the amplitude, $k\in\MR$ is the wave vector, and $\om\in\MR$ is
the frequency of the wave. Then wave equation (\ref{qnbhfy}) in cylindrical
coordinates reduces to the Bessel equation
\begin{equation}                                                  \label{qmmoip}
  r^2\frac{d^2U}{dr^2}+r\frac{dU}{dr}+(\kappa^2r^2-1)U=0,
\end{equation}
where
\begin{equation}                                                  \label{qabspp}
  \kappa^2:=\frac{\om^2}{c_\St^2}-k^2,\qquad c_\St^2:=\frac\mu{\rho_0}.
\end{equation}
A general solution of this equation depends on two integration constants.
Therefore general solutions for inner and outer tubes are
\begin{equation}                                                  \label{qbvxgo}
  W=\begin{cases}
    W_\text{in}=C_1J_1(\kappa y)+C_2N_1(\kappa y), & r_0\le y\le r_1, \\
    W_\text{ex}=C_3J_1(\kappa y)+C_4N_1(\kappa y), & r_2\le y\le r_3,
\end{cases}
\end{equation}
where $J_1$ is the Bessel function of the first kind and first order, $N_1$ is
the Neumann function of the first order (see, i.e.\ \cite{JaEmLo60}), and
$C_{1,2,3,4}$ are integration constants.

To find the integration constants we impose boundary conditions. The boundary
surfaces are assumed to be free, i.e.\ the deformation tensor on the boundary
must be zero
\begin{equation*}
  \left.\frac{dW_\text{in}}{dr}\right|_{r=r_\text{in}}=0,\qquad
  \left.\frac{dW_\text{ex}}{dr}\right|_{r=r_\text{ex}}=0.
\end{equation*}
Because
\begin{equation*}
  \frac{dW}{dr}=\frac{dy}{dr}\frac{dW}{dy}=(1-v)\frac{dW}{dy},
\end{equation*}
and $v(r_\text{in})=v(r_\text{ex})=0$, these equalities in the initial
coordinates take the form
\begin{align*}
  C_1J'_1(z_0)+C_2N'_1(z_0)&=0
\\
  C_3J'_1(z_3)+C_4N'_1(z_3)&=0,
\end{align*}
where
\begin{equation*}
  z:=\kappa y,
\end{equation*}
and prime denotes differentiation with respect to the argument $z$. These
equalities define two integration constants:
\begin{equation}                                                  \label{qbvvgc}
\begin{split}
  C_2&=-k_0C_1,\qquad k_0:=\frac{J'_1(z_0)}{N'_1(z_0)},
\\
  C_4&=-k_3C_1,\qquad k_3:=\frac{J'_1(z_3)}{N'_1(z_3)}.
\end{split}
\end{equation}

On the gluing surface, we impose two conditions: entirety and equality of
stresses,
\begin{equation}                                                  \label{qbvxgl}
  W_\text{in}(r_*)=W_\text{ex}(r_*),\qquad
  \left.\frac{dW_\text{in}}{dr}\right|_{r=r_*}
  \left.\frac{dW_\text{ex}}{dr}\right|_{r=r_*}.
\end{equation}
As a result we get two equations
\begin{equation}                                                  \label{qnvbgt}
\begin{split}
  C_1\big[J_1(z_1)-k_0N_1(z_1)\big]-C_3\big[J_1(z_2)-k_3N_1(z_2)\big]=0,
\\
  C_1\big[J'_1(z_1)-k_0N'_1(z_1)\big]-C_3\big[J'_1(z_2)-k_3N'_1(z_2)\big]=0.
\end{split}
\end{equation}
The necessary and sufficient condition for this system to have a nontrivial
solution is the equality of its determinant to zero:
\begin{multline}                                                  \label{qbsioo}
  \big[J_1(z_1)-k_0N_1(z_1)\big]\big[J'_1(z_2)-k_3N'_1(z_2)\big]-
\\
  -\big[J_1(z_2)-k_3N_1(z_2)\big]\big[J'_1(z_1)-k_0N'_1(z_1)\big]=0.
\end{multline}
For given parameters of the double wall tube $r_0$, $r_1$, $r_2$ и $r_3$, the
obtained relation is the equation for the constant $\kappa$. Let $\kappa$ be a
root of Eq. (\ref{qbsioo}), then the equality
\begin{equation}                                                  \label{qnjuxp}
  \om=c_\St\sqrt{k^2+\kappa^2}.
\end{equation}
defines the dispersion relation.

The phase velocity of torsional waves $v:=\om/k$ is easily found from dispersion
relation (\ref{qnjuxp}):
\begin{equation}                                                  \label{qnjuip}
  v=c_\St\sqrt{1+\frac{\kappa^2}{k^2}}.
\end{equation}
The group velocity is also easily found
\begin{equation}                                                  \label{qnjikp}
  v_{\rm g}:=\frac{d\om}{dk}=\frac{c_\St^2}v.
\end{equation}
We see that the phase velocity of torsional waves is always greater then the
velocity of transverse waves, and group velocity is smaller. Dispersion relation
(\ref{qnjuxp}) does depend on double wall parameters through Eq.(\ref{qbsioo}).
\section{Conclusion}
We found the induced metric in double wall tube with cylindrical dislocation in
the framework of the geometrical theory of defect. Though components of this
metric are not continuous functions, the three dimensional Einstein equations
are well defined (\cite{deBKat09}). Afterwards propagation of torsional waves in
double wall tube is described. The presence of the cylindrical dislocation
inside the double wall tube leads to changing of the dispersion relation.

Double wall tube may be useful as a continuous model of double wall nanotubes.

This work was supported by the Russian Science Foundation (project 14-11-00687)
in Steklov Mathematical Institute.

\end{document}